\documentclass[%
 reprint,
superscriptaddress,
 amsmath,amssymb,
 aps,floatfix
]{revtex4-2}

\usepackage{graphicx}
\usepackage{dcolumn}
\usepackage{placeins}
\usepackage{bm}
\usepackage{svg}
\usepackage{gensymb}
\usepackage{float}

\usepackage{caption}
\usepackage{braket}
\usepackage{xcolor}

\usepackage{siunitx}

\begin{document}

\title{Sliding Luttinger Liquid and Topological Flat Bands \\in Symmetry Mismatched Moir\'e Interfaces}

\author{Abhijat Sarma}
\affiliation{%
 Department of Physics, Cornell University, Ithaca, New York 14853, USA 
}
\affiliation{
 Department of Physics, University of California, Santa Barbara, California 93106, USA
}

\begin{abstract}
In this work we analyze a class of Moir\'e models consisting of an active honeycomb monolayer such as graphene or a hexagonal transition-metal dichalcogenide (TMD) on top of a substrate, in which the K and K' valleys of the active layer are folded near each other by a suitably chosen substrate geometry. Generalizing the so-called ``coupled-valley'' model of Ref.~\cite{Kagome_flat_bands}, we start from a microscopic tight-binding description, deriving a continuum model from Schrieffer-Wolff perturbation theory and obtaining an effective description of the low-energy momentum states in either valley as well as the explicit microscopic forms of the Moir\'e potentials. We then consider two explicit symmetry-mismatched Moir\'e geometries with a rectangular substrate, the first of which displays an emergent time-reversal symmetry as well as a broad parameter regime which displays quasi-1D physics characterized by the existence of a Sliding Luttinger Liquid phase. This model also has a nontrivial topological character, captured by the Berry curvature dipole. The second geometry displays an emergent $C_3$ rotational symmetry despite the rectangular substrate, reducing to a continuum model considered in Ref.~\cite{Kagome_flat_bands} that was shown to display honeycomb and Kagome topological flat bands. 

\end{abstract}

\maketitle

\section{Introduction}

Recently, Moir\'e materials have emerged as a highly tunable platform for realizing strongly correlated physics. In particular, various graphene and TMD based Moir\'e systems have displayed strongly correlated topological flat bands \cite{bm_graphene, tbg_topological, Kagome_flat_bands, Kekule_graphene, Hypermagic_flat_bands, gamma_valley_tmd, Tmd_top_ins, moire_cfl} which host a wide class of interesting strongly interacting phases, such as correlated insulators, topological insulators, integer and fractional quantum anomalous hall states, composite Fermi liquids, and unconventional superconductivity \cite{tbg_correlated_insulator, Tmd_top_ins, tbg_tunable, moire_qah, mote2_fqah, moire_cfl, tbg_sup, tbg_sup_2}. Another interesting class of strongly interacting systems are non-Fermi liquids, such as the Luttinger liquid describing a one dimensional electron gas \cite{luttinger_ll, haldane_ll, tomonoga_ll}, which displays spin-charge separation and a diverging zero-temperature conducivity \cite{ll_cond}. Two dimensional systems with significant anisotropy, i.e. in the case where band dispersion is heavily suppressed along one direction, can display a similar quasi-1D phase known as a Sliding Luttinger Liquid (SLL) \cite{sll_theory, xgwen_sll, ashvin_sll, sll_exp, sll_exp_2}, in which the transverse conductivity obeys a power law and the longitudinal conductivity behaves like that of a Luttinger Liquid. This phase describes the physics of weakly-coupled 1D wires \cite{sll_theory}. Recent experiments and theoretical models have shown Sliding Luttinger Liquid behavior in various Moir\'e materials, such as twisted bilayer WTe$_2$ and black phosphorous \cite{Coupled_wire, wte2_sll_theory, WTe2_sll_exp, anisotropy-moire}. In these settings, the Moir\'e lattice strongly enhances the underlying anisotropy of the monolayer. 

Recently, several Moir\'e heterostructures have been proposed where two inequivalent valleys in a single layer are van der Waals coupled by a substrate, as opposed to the more usual setting in which the same valley of two different layers are coupled together \cite{Kagome_flat_bands, Kekule_graphene, coupled_valley_2, coupled_valley_3, coupled_valley_4, coupled_valley_5}. Such configurations are interesting because there is more freedom in choosing the substrate, which need not even have the same lattice type as the active layer, allowing one to construct Moir\'e lattices of different symmetries from the same active layer geometry. In this work, we construct the general ``coupled-valley'' continuum model describing the setting in which the K and K' valleys of a honeycomb active layer are coupled by a suitably chosen substrate. Starting from a microscopic tight-binding description, we derive the general form of the Moir\'e potentials under the two-center approximation. We then apply our theory to two ``symmetry-mismatched'' geometries in which the honeycomb layer is placed on a rectangular substrate. The first possesses a rectangular Moir\'e unit cell, and displays Sliding Luttinger Liquid physics, as well as a nonzero Berry curvature dipole, owing to the broken $C_3$ rotational symmetry of the Moir\'e model. The second possesses a hexagonal Moir\'e unit cell, and the continuum model has an effective $C_3$ rotational symmetry despite the substrate having no such rotational symmetry. This model reduces to one previously studied in Ref.~\cite{Kagome_flat_bands}, which was shown to display topological flat bands that are in agreement with Kagome and honeycomb tight-binding models on the Moir\'e lattice. 

\section{General Coupled-Valley Continuum Model}

In this section, we consider a generic geometry giving rise to a coupled-valley model, and present our results for the continuum model derived using the Schrieffer-Wolff perturbation theory. The details of the derivation can be found in Appendix \ref{derivation}. In this paper, we use the notation $\mathrm{Span}(\{\mathbf{v}_i\})$ to denote the set of linear combinations of $\{\mathbf{v}_i\}$ with integer coefficients. We assume a Moir\'e bilayer geometry, where the top (active) layer is a honeycomb lattice whose low-energy excitations are described by Dirac cones which may or may not be massive lying at the K and K' valleys. We denote the primitive reciprocal lattice vectors in the active layer by $\mathbf{b}_1^+, \mathbf{b}_2^+ = \frac{4\pi}{a\sqrt{3}}(-\frac{\sqrt{3}}{2}, \pm \frac{1}{2})$. The K and K' valleys are then centered at $\pm \mathbf{K}^+ = \mp \frac{1}{3}(\mathbf{b}_1^+ + \mathbf{b}_2^+)$. We also denote the orbital positions in the unit cell by $\boldsymbol\tau_{1}^+, \boldsymbol\tau_{2}^+ = a(0, \pm\frac{1}{2\sqrt{3}})$. We assume that the active layer possesses time reversal symmetry and that the valleys are time reversal conjugates of one-another, so that the isolated active layer Hamiltonian acting on the two valleys takes the form $H_0(\mathbf{k}) = \begin{bmatrix}
\hbar v_{F} \boldsymbol{\sigma}\cdot \mathbf{k} + \frac{m}{2}\sigma_{z} & 0\\
0 & -\hbar v_{F} \boldsymbol{\sigma}^*\cdot \mathbf{k} + \frac{m}{2}\sigma_{z}
\end{bmatrix}$, where $\boldsymbol{\sigma} = (\sigma_x, \sigma_y)$, $^*$ denotes complex conjugation, and the $\sigma$'s act on the orbital index. The bottom (substrate) layer lattice vectors and orbital positions $\mathbf{b}_1^-, \mathbf{b}_2^-, \{\boldsymbol\tau_{i}^-\}$ will be left unspecified for now. In particular, we make no assumptions on the geometry of the substrate lattice, which need not be triangular, nor need it share any discrete symmetries with the active layer. We account for the Moir\'e lattice mismatch by writing $\mathbf{b}_{i}^-= R_\theta e^{-\epsilon} \mathbf{b}_{i}^0$, $\boldsymbol{\tau}_i^-= R_\theta e^{-\epsilon} \boldsymbol{\tau}_i^0$, where $R$ is a rotation matrix, $\theta$ and $\epsilon$ are small parameters, and $\mathbf{b}_{1}^0, \mathbf{b}_{2}^0$ are the reciprocal vectors corresponding to a \textit{commensurate} substrate geometry, i.e. a geometry in which the two layers form a commensuration superlattice, or equivalently, $L^c = \mathrm{Span}(\mathbf{b}_{1}^+, \mathbf{b}_{2}^+) \cap \mathrm{Span}(\mathbf{b}_{1}^0, \mathbf{b}_{2}^0)$ contains more than just the zero vector. If the set is nonempty, then it can be written as $L^c = \mathrm{Span}(\mathbf{b}_1^c, \mathbf{b}_2^c)$ where $\mathbf{b}_1^c, \mathbf{b}_2^c$ are linearly independent and satisfy linear equations $\mathbf{b}_i^c = \sum_{j}A_{ij}\mathbf{b}_j^+ = \sum_{j}B_{ij}\mathbf{b}_j^0$ where $A_{ij}, B_{ij}$ are integers (such a choice of primitive vectors is nonunique). We then define the Moir\'e reciprocal lattice primitive vectors $\mathbf{b}_i^M = (\sum_{j}A_{ij}\mathbf{b}_j^+) - (\sum_{j}B_{ij}\mathbf{b}_j^-) = (1-R_\theta e^{-\epsilon})\mathbf{b}_i^c$.

The effect of the substrate layer will be to induce a Moir\'e potential, allowing for electron hopping between the two layers. The substrate geometry will be chosen such that this coupling allows for second order hopping processes $\pm \mathbf{K} + \mathbf{k} \rightarrow \mathrm{Substrate} \rightarrow \mp \mathbf{K} + \mathbf{k'}$, where $\mathbf{k}$ and $\mathbf{k'}$ are both small (i.e., on the Moir\'e scale). This will lead to mixing between the K and K' valley Dirac cones, allowing for interesting physics to emerge on the Moir\'e scale. It is a sufficient condition to get a coupled-valley model to require only that the set of k-points $+\tilde{S}^0 = (\mathbf{K^+}+\mathrm{Span}(\mathbf{b}_1^+, \mathbf{b}_2^+, \mathbf{b}_1^0, \mathbf{b}_2^0)) \cap BZ^0$ has at least one element $\tilde{\mathbf{s}}^0 \in +\tilde{S}^0$ such that $-\tilde{\mathbf{s}}^0 \in (\tilde{\mathbf{s}}^0+\mathrm{Span}(\mathbf{b}_1^0,\mathbf{b}_2^0))$, constraining the set of allowed substrate geometries. We denote the set of elements of $+\tilde{S}^0$ which satisfy this condition as $S^0$, and its elements by $\mathbf{s}_i^0$. 

To each $\mathbf{s}_i^0$, we associate another k-point $\mathbf{w}_i^0$ which is a minimal norm element of $(\mathbf{s}_i^0+\mathrm{Span}(\mathbf{b}_1^0, \mathbf{b}_2^0))\cap(\mathbf{K^+}+\mathrm{Span}(\mathbf{b}_1^+, \mathbf{b}_2^+))$. We also define the incommensurate versions of these momenta which appear in the Moir\'e geometry, $\mathbf{s}_i^- =R_{\theta} e^{-\epsilon} \mathbf{s}_i^0$, $\mathbf{w}_i^- =R_{\theta} e^{-\epsilon} \mathbf{w}_i^0$. We assume that the substrate Hamiltonian at $\mathbf{s}_i^-$, $H_i^- = H^-(\mathbf{s}_i^-)$ is gapped relative to the active Fermi level. Lastly, we define $\mathbf{q}_i = \mathbf{w}_i^- - \mathbf{w}_i^0$. It is then straightforward to write down the effective low-energy Hamiltonian for the active layer using Schrieffer-Wolff perturbation theory. We define creation operators $a^\dag_{\mathbf{k}}, b^\dag_{\mathbf{k}}$ corresponding to creating Bloch states in each valley, i.e $a^\dag_{\mathbf{k}} = c^\dag_{\mathbf{K^+}+\mathbf{k}}$, $b^\dag_{\mathbf{k}} = c^\dag_{-\mathbf{K^+}+\mathbf{k}}$ where $c^\dag_{\mathbf{k}}$ is the usual creation operator for a Bloch state in the active layer and we have suppressed the spin index. Here we have collected the orbital degree of freedom into a vector, i.e. $c^\dag_{\mathbf{k}} = (c^\dag_{\mathbf{k}, A}, c^\dag_{\mathbf{k}, B})$ where $A/B$ denotes the sublattice. For shorthand, let $\mathbf{k}_{nm} = \mathbf{k}+n\mathbf{b}_1^M+m\mathbf{b}_2^M$. Then, the spinless single-particle Hamiltonian is given in second-quantized form by 

\begin{align}
\begin{split}
\label{general_cont_model}
    &\hspace{2cm}H = \frac{\Omega_M}{(2\pi)^2}\int_{MBZ} d^2\mathbf{k} H(\mathbf{k}) \\&\\
&H(\mathbf{k}) = \sum_{n,m} [a^\dag_{\mathbf{k}_{nm}} (\hbar v_{F}\boldsymbol{\sigma}\cdot\mathbf{k}_{nm}+\frac{m}{2}\sigma_z) a_{\mathbf{k}_{nm}} \\
& + \sum_{i}b^\dag_{\mathbf{k}_{nm}+2\mathbf{q}_i}(-\hbar v_{F}\boldsymbol{\sigma}^*\cdot(\mathbf{k}_{nm}+2\mathbf{q}_i)+\frac{m}{2}\sigma_z)  b_{\mathbf{k}_{nm}+2\mathbf{q}_i}] \\
&+ \sum_{n_1,m_1,n_2,m_2} \{a^\dag_{\mathbf{k}_{n_2 m_2}}S_{(n_2-n_1),(m_2-m_1),+}a_{\mathbf{k}_{n_1 m_1}} \\
&+\sum_{i}[b^\dag_{\mathbf{k}_{n_1 m_1}+2\mathbf{q}_i}S_{(n_2-n_1),(m_2-m_1),-}b_{\mathbf{k}_{n_2 m_2}+2\mathbf{q}_i} \\
&+ b^\dag_{\mathbf{k}_{n_2 m_2}+2\mathbf{q}_i}T_{(n_2-n_1),(m_2-m_1),i}a_{\mathbf{k}_{n_1 m_1}} + \mathrm{h.c.}]\}
\end{split}
\end{align}

\noindent where 

\begin{align}
\begin{split}
\label{intravalley_pot}
&S^{\alpha\beta}_{n,m,\pm} = \frac{-1}{|\Omega_+||\Omega_-|}\sum_{i,n_1,m_1,\gamma,\gamma'} \hat{t}^{\alpha\gamma}(\pm\mathbf{w}_i^0\pm n_1\mathbf{b}_1^c\pm m_1\mathbf{b}_2^c) 
\\&((H_i^-)^{-1})^{\gamma \gamma'}(\hat{t}^{\beta\gamma'}(\pm\mathbf{w}_i^0\pm(n_1-n)\mathbf{b}_1^c\pm(m_1-m)\mathbf{b}_2^c))^* 
\\&e^{\pm i(\phi_{n_1 m_1 i}^{\alpha\gamma} - \phi_{(n-n_1)(m-m_1) i}^{\beta\gamma'})}
\end{split}
\end{align}

\noindent and 

\begin{align}
\begin{split}
\label{intervalley_pot}
&T^{\alpha\beta}_{n,m,i} = \frac{-1}{|\Omega_+||\Omega_-|}\sum_{n_1,m_1,\gamma,\gamma'} \hat{t}^{\alpha\gamma}(\mathbf{w}_i^0+n_1\mathbf{b}_1^c+m_1\mathbf{b}_2^c) 
\\&((H_i^-)^{-1})^{\gamma \gamma'}(\hat{t}^{\beta\gamma'}(-\mathbf{w}_i^0+(n_1-n)\mathbf{b}_1^c+(m_1-m)\mathbf{b}_2^c))^* 
\\&e^{i(\phi_{n_1 m_1 i}^{\alpha\gamma} + \phi_{(n-n_1)(m-m_1) i}^{\beta\gamma'})}
\end{split}
\end{align}

\noindent are the intravalley and intervalley tunnelings, respectively. In these equations, $\hat{t}^{\alpha \beta} (\mathbf{k})$ is the momentum-space hopping amplitude (which can be calculated microscopically from DFT, for example) and $$e^{i\phi_{nmi}^{\alpha\beta}} = e^{i(n\mathbf{b}_1^c+m\mathbf{b}_1^c)\cdot(\boldsymbol{\tau}_{\alpha}^+-\boldsymbol{\tau}_{\beta}^0)} e^{i[(\mathbf{w}_i^0-\mathbf{K^+})\cdot\boldsymbol{\tau}_{\alpha}^+ - (\mathbf{w}_i^0-\mathbf{s}_i^0)\cdot\boldsymbol{\tau}_{\beta}^0)]}$$ is a geometric phase associated with the hoppings. In practice, $\hat{t}^{\alpha \beta} (\mathbf{k})$ decays quickly, and only the first few tunneling matrices in the Hamiltonian need to be kept.

\section{Anisotropy Model}
\label{sec:anisotropy_model}

\begin{figure*}[t]
    \centering
    \hspace*{-3cm}\includegraphics[height=10cm, width=22cm]{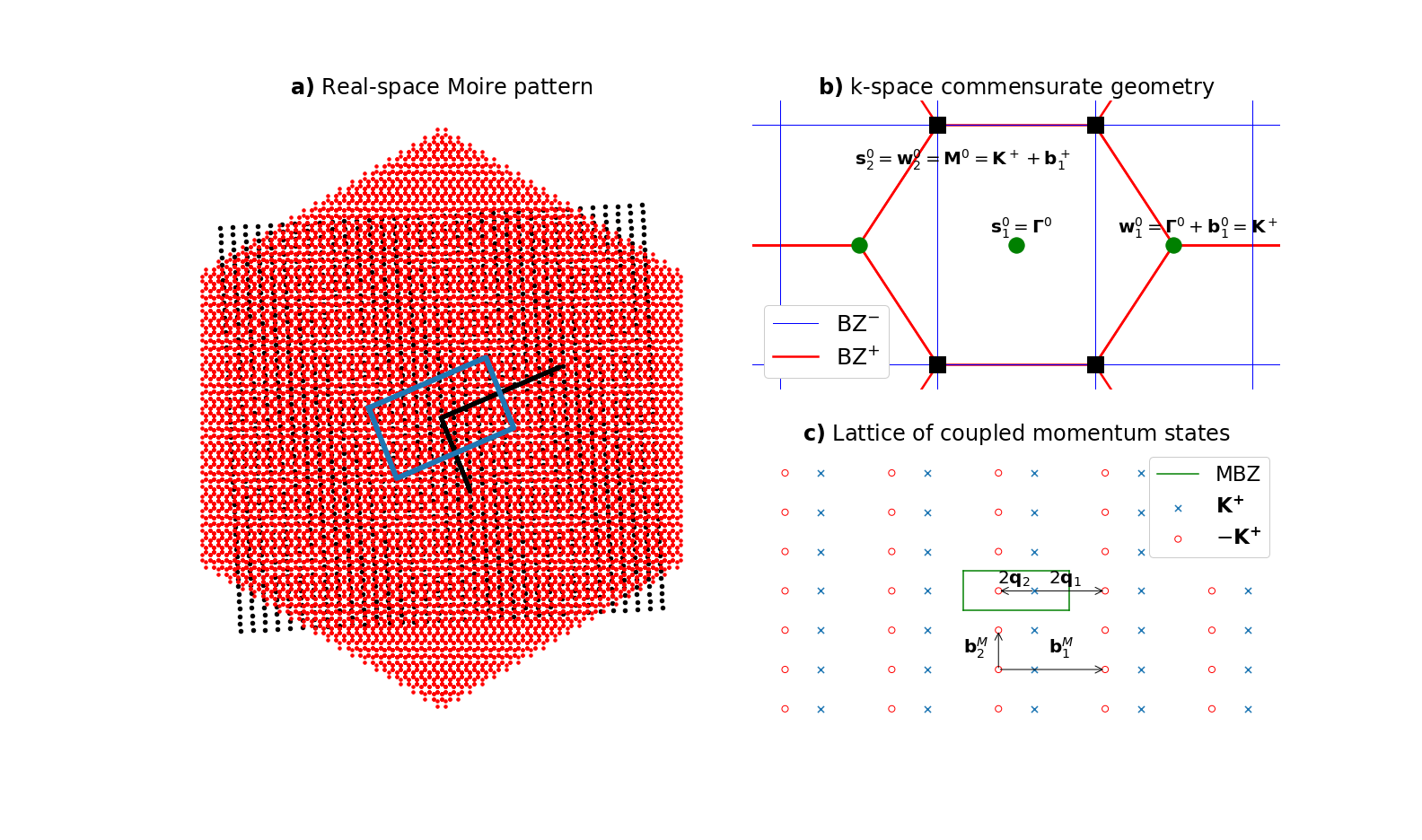}
    \caption{\textbf{Anisotropy Model Geometry}. \textbf{a)} Moir\'e pattern, $\theta=3\degree$ and $\epsilon=0.02$. Blue rectangle denotes Moir\'e unit cell, and black arrows denote Moir\'e primitive vectors. \textbf{b)} Momentum space commensurate geometry. Low-energy Moir\'e couplings are generated by substrate states near $\boldsymbol{\Gamma}^-$ and $\textbf{M}^-$. \textbf{c)} Lattice of momentum states coupled by Moir\'e potential.}
    \label{fig:anisotropy_geometry}
\end{figure*}

For the anisotropy model, we consider the case of an active layer with a mass gap, and a rectangular substrate geometry given by $\mathbf{b}_1^0 = (\frac{4\pi}{3a}, 0)$, $\mathbf{b}_2^0 = (0, \frac{4\pi}{a\sqrt{3}})$, $\mathbf{b}_i^- = R_{\theta}e^{\epsilon}\mathbf{b}_i^0$. The active layer can be a TMD monolayer, or another gapped K-valley material such as hexagonal Boron Nitride (hBN) or a graphene-hBN heterostructure \cite{graphene_hbn}. With this geometry, $\mathbf{b}_i^c = \mathbf{b}_i^0$, and therefore the Moir\'e BZ is rectangular. Further, in this case $S^0 = \{\boldsymbol{\Gamma}^0, \mathbf{M}^0\}$ where $\mathbf{M}^0 = \frac{1}{2}(-\mathbf{b}_1^0+\mathbf{b}_2^0)$. Thus, the TMD states in either valley may tunnel into the substrate layer at k-points near $\{\boldsymbol{\Gamma}^-, \mathbf{M}^-\}$ before tunneling back. When tunneling through $\boldsymbol{\Gamma}^-$, the associated momentum offset from changing valleys is $2\mathbf{q}_1 = \frac{2}{3}\mathbf{b}_1^M$. When tunneling through $\mathbf{M}^-$, the associated momentum offset is $2\mathbf{q}_2 = -\frac{1}{3}\mathbf{b}_1^M$. These two vectors are equivalent modulo the Moir\'e reciprocal lattice, and therefore generate the same lattice of coupled momentum states according to Eq.~\ref{general_cont_model}. This geometry is pictured in Figure \ref{fig:anisotropy_geometry}.

For simplicity, we assume that the substrate layer has time-reversal symmetry, and contains only one active orbital near the active layer fermi energy, located at $\boldsymbol{\tau^-} = (0, 0)$. In the case that the active layer is a TMD material, we must account for the presence of spin-valley locking; namely, the effect that the low-energy modes in the K/K' valleys have opposite spin, and thus cannot be directly coupled by simple tunneling effects. This problem can be circumvented by the presence of an in-plane magnetic field in the substrate layer. We assume the substrate has negligible spin-orbit coupling, so that at the $\boldsymbol{\Gamma}^-$ and $\mathbf{M}^-$ points, the two spinful bands are degenerate. The in-plane magnetic field will then mix the two spin sectors at these points in the substrate, allowing for intervally tunneling through the substrate. For specificity, we consider the substrate Hamiltonian at $\boldsymbol{\Gamma}^-$ and $\mathbf{M}^-$ to be $H_i^- = \begin{bmatrix}
    U_i^- & g\mu_{B}B \\
    g\mu_{B}B & U_i^-
\end{bmatrix}$, where $U_1^-$ is the energy of the substrate electrons at the $\boldsymbol{\Gamma}^-$ point without an applied field relative to the TMD Fermi energy, and $U_2^-$ is the same for the $\mathbf{M}^-$ point. Then, $(H_i^-)^{-1} = \begin{bmatrix}
    \frac{U_i^-}{(U_i^-)^2 - (g\mu_{B}B)^2} & \frac{-g\mu_{B}B}{(U_i^-)^2 - (g\mu_{B}B)^2} \\
    \frac{-g\mu_{B}B}{(U_i^-)^2 - (g\mu_{B}B)^2} & \frac{U_i^-}{(U_i^-)^2 - (g\mu_{B}B)^2}
\end{bmatrix}$. The diagonal elements of this matrix correspond to intravalley hopping, while the off-diagonal elements correspond to intervalley hopping. Thus, the intravalley potentials (Eq.~\ref{intravalley_pot}) scale like $(\hat{t})^2\frac{U^-}{(U^-)^2 - (g\mu_{B}B)^2}$, while the intervalley potentials (Eq.~\ref{intervalley_pot}) scale like $(\hat{t})^2\frac{g\mu_{B}B}{(U^-)^2 - (g\mu_{B}B)^2}$. 

Keeping only the few lowest order tunneling matrices, and defining $\mathbf{k}_{nm}^{\pm} = \mathbf{k}+(n \pm \frac{1}{6})\mathbf{b}_1^M + m\mathbf{b}_2^M$, we can write the Hamiltonian as 

\begin{align}
\begin{split}
\label{anisotropy_model}
&\hspace{2cm}H = \frac{\Omega_M}{(2\pi)^2}\int_{MBZ} d^2\mathbf{k} H(\mathbf{k}) \\&\\
&H(\mathbf{k}) = \sum_{n,m} [a^\dag_{\mathbf{k}_{nm}^+}(\hbar v_{F}\boldsymbol{\sigma}\cdot\mathbf{k}_{nm}^+ +\frac{m}{2}\sigma_z + S_0) a_{\mathbf{k}_{nm}^+} \\
& + b^\dag_{\mathbf{k}_{nm}^-}(-\hbar v_{F}\boldsymbol{\sigma}^*\cdot\mathbf{k}_{nm}^- +\frac{m}{2}\sigma_z + S_0^*)  b_{\mathbf{k}_{nm}^-} \\
&+ a^\dag_{\mathbf{k}_{(n+1)m}^+} S_1 a_{\mathbf{k}_{nm}^+} + a^\dag_{\mathbf{k}_{n(m+1)}^+} S_2 a_{\mathbf{k}_{nm}^+} + \mathrm{h.c.} \\
&+ b^\dag_{\mathbf{k}_{(n-1)m}^-} S_1^* b_{\mathbf{k}_{nm}^-} + b^\dag_{\mathbf{k}_{n(m-1)}^-} S_2^* b_{\mathbf{k}_{nm}^-} + \mathrm{h.c.} \\
&+ b^\dag_{\mathbf{k}_{nm}^-} T_1 a_{\mathbf{k}_{nm}^+} + b^\dag_{\mathbf{k}_{(n+1)m}^-} T_2 a_{\mathbf{k}_{nm}^+} + \mathrm{h.c.} \\
& + b^\dag_{\mathbf{k}_{n(m+1)}^-} T_3 a_{\mathbf{k}_{nm}^+} + b^\dag_{\mathbf{k}_{n(m-1)}^-} T_4 a_{\mathbf{k}_{nm}^+} + \mathrm{h.c.}]
\end{split}
\end{align}

\noindent and it is implicitly understood that $a^\dag$ creates a spin up state while $b^\dag$ creates a spin down state. The Hamiltonian has a mirror symmetry $M_x$ that acts on the Bloch states by $M_x \ket{k_x, k_y, \alpha} = \ket{-k_x, k_y, \alpha}$. Despite the explicit time-reversal symmetry breaking from the magnetic field in the substrate in the case of a TMD active layer, the low-energy Hamiltonian also possesses an effective spinless anti-unitary time-reversal symmetry $\tau$ that satisfies $\tau^2=1$ and acts on the TMD Bloch states by $\tau\ket{k_x, k_y, \alpha} = \ket{-k_x, -k_y, \alpha}$. It therefore acts on the creation operators by $\tau a^\dag_{\mathbf{k}}\tau = b^\dag_{-\mathbf{k}}$. One can straightforwardly verify that $\tau$ is a symmetry of the Hamiltonian by plugging in the microscopic forms of the tunneling matrices (Eq.~\ref{intravalley_pot},~\ref{intervalley_pot}) and showing that $\tau H(\mathbf{-k}) \tau = H(\mathbf{k})^*$, implying that the spectrum of $H(\mathbf{k})$ is the same as that of $H(-\mathbf{k})$. This time-reversal symmetry exists because, to lowest order, the magnetic field on the substrate does not change the interlayer tunneling amplitude; it only serves to mix the two spin sectors in the substrate.

Save for the symmetries enforced by the microscopic form of the tunneling matrices ($M_x$ and $\tau$), we now treat them as arbitrary tuning parameters. Utilizing the symmetry constraints, the $2\times2$ tunneling matrices $S_0, S_1, S_2, T_1, T_2, T_3, T_4$ can be parameterized by 25 real parameters, shown in Appendix \ref{app:anisotropy_params}. Generally speaking, these tunneling matrices have the tendency to generate a band structure which is strongly anisotropic, in the sense that there is a large parameter regime in which the dispersion is strongly suppressed in one direction. 

\begin{figure*}[t]
    \centering
    \hspace*{-.9cm}\includegraphics[height=9cm, width=17cm]{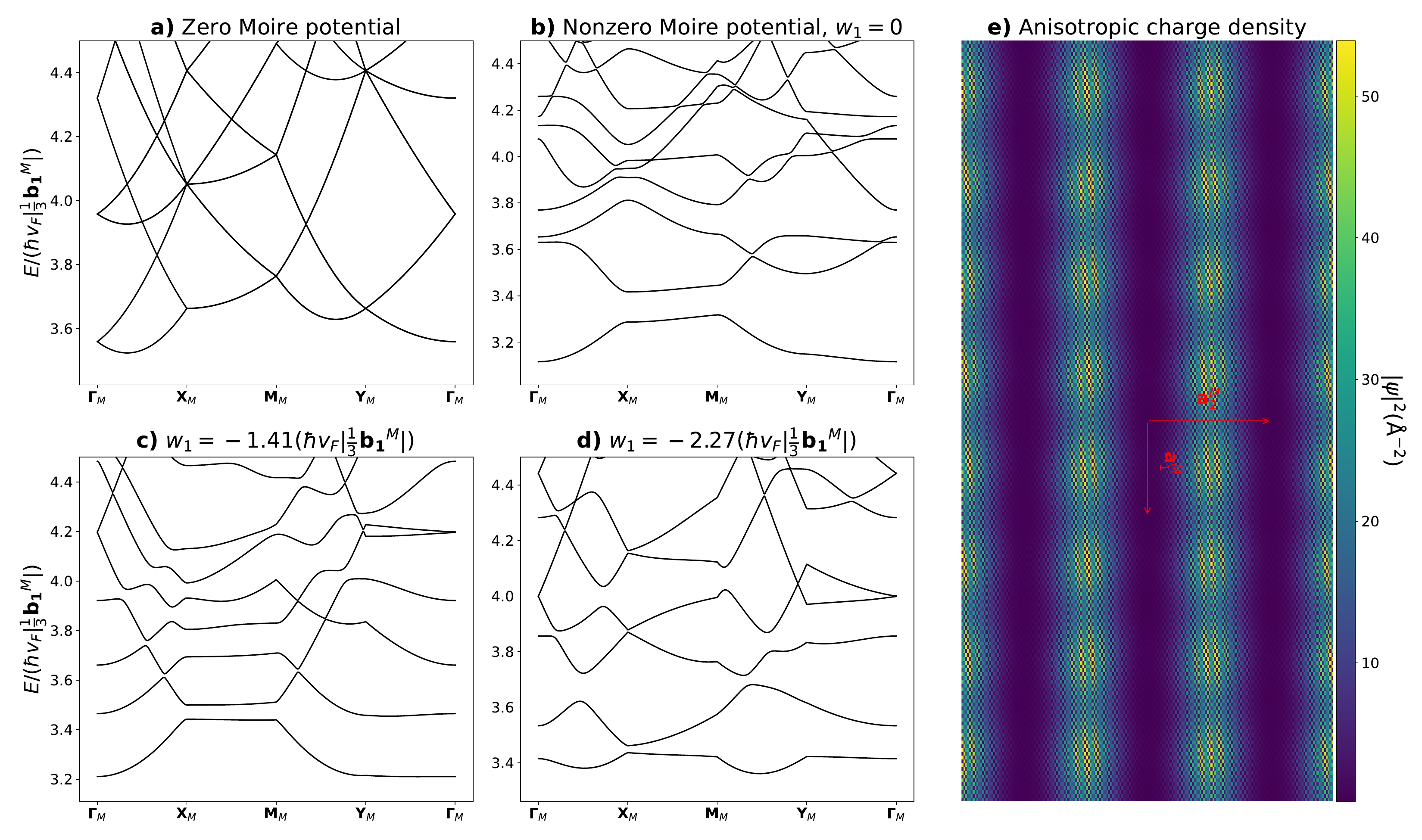}
    \caption{\textbf{Band Anisotropy Origin From Moir\'e Potential}. \textbf{a)}--\textbf{d)} Evolution of the band structure as the Moir\'e potential is turned on. Only conduction bands are shown. In \textbf{b)}--\textbf{d)}, the hopping terms ($S_1, S_2, T_{1-4}$) are all nonzero and on the order of $50\mathrm{meV}$, while the renormalizing term $w_1$ varies. Even when $w_1=0$, there is some anisotropy, as the bandwidth along the y-direction is roughly half the bandwidth along the x-direction. As $w_1$ increases, this anisotropy maximizes in \textbf{c)}, where the band almost perfectly flattens along the y-direction. \textbf{e)} Charge density from fully filling only the first conduction band in \textbf{c)}. Wires of high charge density are clearly visible. Here $\mathbf{a}_i^M$ are the real-space Moir\'e lattice vectors.}
    \label{fig:anisotropy_bs}
\end{figure*}

For the sake of explaining this effect, we note the explicit form of $S_0$, given by $S_0 = w_0 \sigma_0 + w_1 \sigma_x + w_2 \sigma_z$. Henceforth we refer to $S_0$ as the renormalizing term, because it renormalizes the bare intravalley dispersion, and the rest of the tunneling matrices as the hopping terms, as they mix different momentum states. The explicit form of the hopping terms are inconsequential, so long as they are nonzero. The hopping terms have the effect of raising the degeneracies of the Moir\'e minibands, opening up gaps. Importantly, when the Moir\'e potential is zero, the lowest energy degeneracy point is $\boldsymbol{\Gamma}^M$, as it is the closest degeneracy point to the origins of the Dirac cones in each valley. Thus, when a gap opens up at this point due to nonzero $T_1$, the energy of the first valence/conduction band can be pushed higher/lower at $\boldsymbol{\Gamma}^M$ ($\mathbf{k}=0$) than at the Dirac cone origins in each valley ($\mathbf{k}=\pm\frac{1}{6}\mathbf{b}_1^M$) due to intervalley mixing, thus pushing the valence band maximum / conduction band minimum (VBM/CBM) to $\boldsymbol{\Gamma}^M$. Once this happens, there is generically some anisotropy, because $\boldsymbol{\Gamma}^M$ is shifted from the Dirac cone origins along the x-direction but not along the y-direction. 

This effect can be strongly enhanced by the renormalizing term, in particular $w_1$, which effectively translates the Dirac cones in each valley in opposite directions along the x-direction. If $|w_1|$ is too large, then the Dirac cone origins will be ``pushed'' into the neighboring Moir\'e Brillouin zones, and the VBM/CBM will move away from the $\boldsymbol{\Gamma}^M$ point. There is generically a sweet-spot of tuning $w_1$ and $|T_1|$ which maximizes the anisotropy. 

These results are displayed in our numerical simulations, which show that there is a ``magic manifold'' of realistic parameter values in which the anisotropy is maximal, around which we predict the existence of a Sliding Luttinger Liquid phase. For our numerical simulations, we choose the Dirac mass as $m=1545 \mathrm{meV}$, and the Dirac velocity as $\hbar v_F = 3949 \mathrm{meV\cdot\r{A}}$, which are the Dirac parameters of $\mathrm{WSe}_2$ \cite{TMD_tb}. We identify $E_{\mathrm{scale}}=\hbar v_F (\frac{1}{3}|\mathbf{b}_1^M|)$ as a natural energy scale of the problem, and choose a Moir\'e configuration in which $E_{\mathrm{scale}}=219 \mathrm{meV}$, corresponding to a Moir\'e lattice vector on the order of $10 \mathrm{\r{A}}$. This energy scale can easily be modified for a given substrate by altering the twisting angle, effectively allowing one to tune the strength of the Moir\'e potentials relative to the monolayer kinetic energy. Figure \ref{fig:anisotropy_bs} demonstrates the evolution of the band structure as the Moir\'e potential is turned on and $|w_1|$ is increased, showing a sweet-spot in which the band is nearly perfectly flat along the $y$ direction. We find that the anisotropy is made largest when $w_1<0$, though strongly anisotropic configurations with $w_1>0$ exist as well. The strongly anisotropic configurations also display a real-space charge density consisting of wires of high charge density separated by regions of nearly zero charge density, further supporting the quasi-one-dimensional character of the band structure. The full list of parameter values used to generate the plots in Figures \ref{fig:anisotropy_bs} and \ref{fig:anisotropy_fs} can be found in Appendix \ref{app:anisotropy_params}.

\begin{figure*}[t]
    \centering
    \hspace*{-.7cm}\includegraphics[height=14cm, width=17cm]{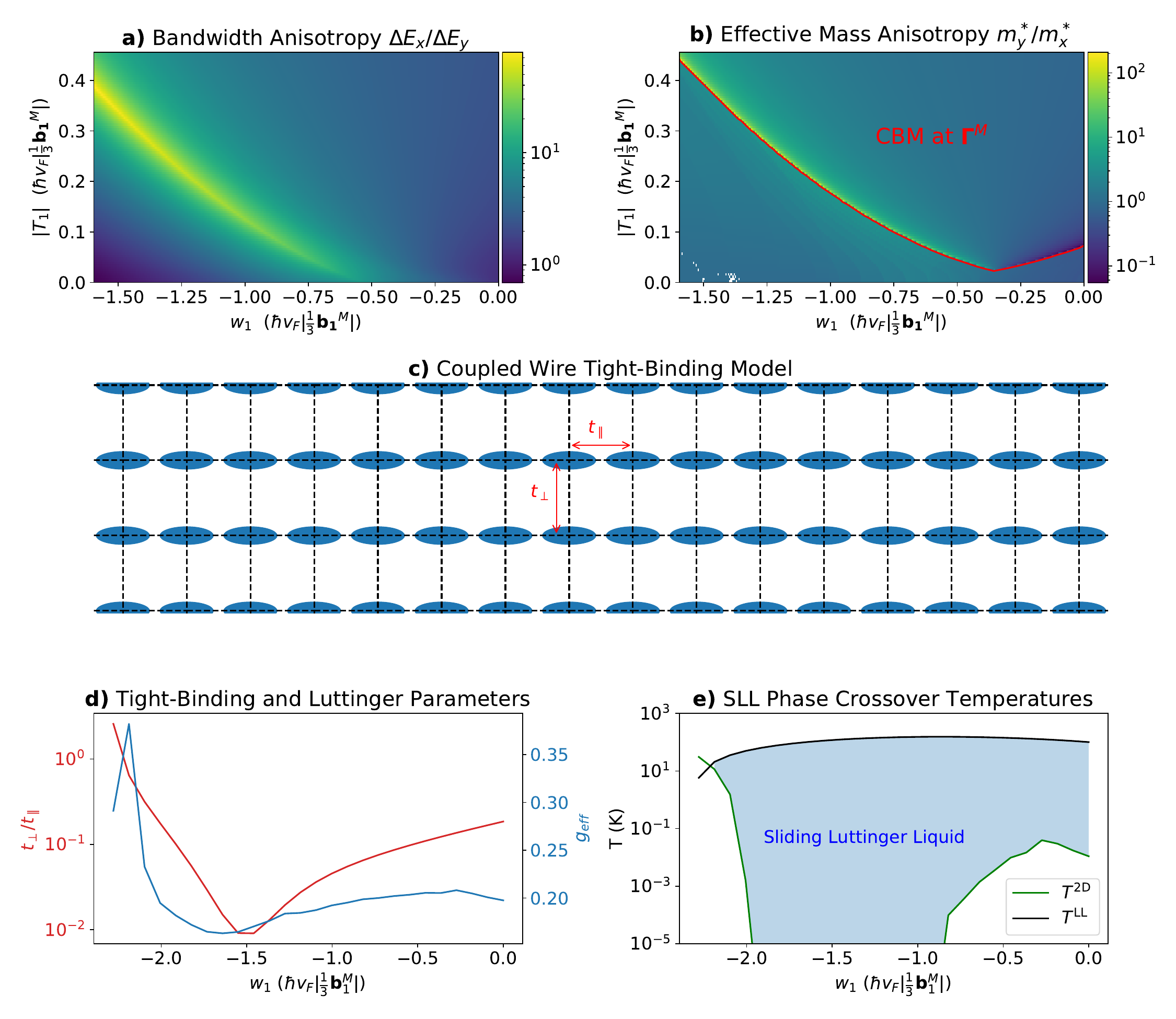}
    \caption{\textbf{Magic Manifold of Anisotropy}. \textbf{a)} Bandwidth anisotropy. Here, the bandwidths in each direction $\Delta E_x$ and $\Delta E_y$ are defined as the width of the first conduction band along the lines $\boldsymbol{\Gamma}^M$ -- $\textbf{X}^M$ and $\boldsymbol{\Gamma}^M$ -- $\textbf{Y}^M$ respectively. \textbf{b)} Effective mass anisotropy. The MBZ is first sampled to find the conduction band minimum (CBM), about which the effective mass is computed in each direction by calculating a quadratic fit to the dispersion in small region about the CBM. Missing points (white) are configurations where the calculation failed to converge to the CBM. \textbf{c)} Quasi-1D tight binding model. Shaded ellipses represent the Wannier centers of the first conduction band (see Figure \ref{fig:anisotropy_bs}\textbf{e}). Dashed lines denote the Moir\'e unit cell. Intrawire and interwire hoppings are represented by $t_{\parallel}, t_{\perp}$. \textbf{d)} Tight binding parameters ratio (red) and Luttinger parameter (blue). \textbf{e)} Sliding Luttinger Liquid phase crossover temperatures.}
    \label{fig:anisotropy_fs}
\end{figure*}

In Figure \ref{fig:anisotropy_fs}, we more thoroughly explore the parameter space and associated anisotropy of the first conduction band. As stated earlier, these simulations demonstrate a ``magic line'' of parameters in which the bandwidth and effective mass ratios of the first conduction band are maximized, roughly corresponding to the regions where the CBM begins to move away from the $\boldsymbol{\Gamma}^M$ point. In other words, the anisotropy is maximized when $|w_1|$ is made as large as possible without pushing the CBM away from $\boldsymbol{\Gamma}^M$, as expected (see Figures \ref{fig:anisotropy_fs}\textbf{a} and \ref{fig:anisotropy_fs}\textbf{b}). When all parameters are allowed to vary, this becomes a ``magic manifold'' of strong anisotropy. 

Around this magic manifold, we can approximate the first conduction (or valence) band by a single-orbital ``coupled-wire'' tight-binding model on the Moir\'e lattice, with nearest-neighbor tunnelings $t_{\parallel}, t_{\perp}$ corresponding to hopping in the x-direction (along the wire) and y-direction (across wires) respectively (Figure \ref{fig:anisotropy_fs}\textbf{c)}). The hopping amplitudes can be calculated by projecting the Hamiltonian to the Wannier basis, giving $t_{\parallel/\perp} = \frac{\Omega_M}{(2\pi)^2}\int_{MBZ} d^2 \mathbf{k} E_{\mathbf{k}}e^{-i \mathbf{k}\cdot \mathbf{a}_{1/2}^M}$ where $E_{\mathbf{k}}$ is the energy of the first conduction band. We also can estimate the effective Luttinger parameter in each wire, given by $g_{\mathrm{eff}} = \mathrm{mean}_{\mathbf{k}\in \mathrm{FS}} (\sqrt{\frac{2\pi\hbar v + \tilde{V}(2k_x)}{2\pi\hbar v + 2 \tilde{V}(0) - \tilde{V}(2k_x)}})$ \cite{Coupled_wire}, where $\tilde{V}(\mathbf{q})$ is the Fourier transform of the interaction potential, $v$ is the Fermi velocity at $\mathbf{k}$, and $\mathrm{mean}_{\mathbf{k}\in \mathrm{FS}}$ denotes averaging over the Fermi surface. This formula arises because for a given wavevector $\mathbf{k}$, $V(0)$ $(V(2k_x))$ is the interaction strength for forward (back) scattering along the wire. As in Ref.~\cite{Coupled_wire}, we consider a screened Coulomb potential of the form $V(r) = U_{I}e^{-r/r_0}$, with strength $U_I=500 \mathrm{meV}$ and screening length $r_0 = 100\mathrm{\r{A}}$ in our simulations, as well as a Fermi energy of $E_F=10\mathrm{meV}$ relative to the bottom of the conduction band.

When $t_{\perp}/t_{\parallel}\ll 1$, $t_{\perp}$ can be treated as a perturbative coupling between 1D wires, which are each described by a Luttinger Liquid. These Luttinger Liquids do not exhibit spin-charge separation due to the spin-orbit coupling \cite{ll_soc}. Incorporating the interwire coupling as a self-energy correction to the 1D Luttinger Liquid Green's function, one finds a divergence when the energy scale $\omega < \tilde{t}_\perp (t_\perp / t_\parallel)^{\frac{\eta}{1-\eta}}$ where $\eta = \frac{1}{4}(g_{\mathrm{eff}} + 1/g_{\mathrm{eff}} - 2)$, giving a lower-bound crossover temperature for the Sliding Luttinger Liquid phase of $T^{\mathrm{2D}} \approx k_B^{-1}t_\perp (t_\perp / t_\parallel)^{\frac{\eta}{1-\eta}}$. This is only the case when $0\leq \eta<1$; if $\eta \geq 1$, then no such divergence exists, and the SLL phase is stable at 0K. Below the crossover temperature, the interwire coupling destroys the Luttinger Liquid, and one generically expects to find a Fermi Liquid or a symmetry broken phase such as a superconductor or a CDW \cite{Coupled_wire, anisotropy-moire}.

Above $T^{\mathrm{2D}}$, the longitudinal conductivity $\sigma_{\parallel}$ is that of a Luttinger Liquid, while the transverse conductivity can be calculated from the Kubo formula as 
$\sigma_{\perp}(V, T) \propto \begin{pmatrix} V^{2\eta-1}, & eV > k_B T\\ T^{2\eta-1}, & eV < k_B T\end{pmatrix}$, where $V$ is the voltage between two neighboring wires and $T$ is the temperature. This behavior cannot persist above the temperature $T^{\mathrm{LL}} \approx k_B^{-1} t_{\parallel}$, above which the intrawire Luttinger liquid is destroyed by thermal fluctuations. Figures \ref{fig:anisotropy_fs}\textbf{d} and \ref{fig:anisotropy_fs}\textbf{e} show $t_\perp / t_\parallel$, $g_{\mathrm{eff}}$, and the crossover temperatures around the magic manifold. They display extended temperature regimes which exhibit the Sliding Luttinger Liquid phases, increasing in size near the magic manifold.

Lastly, the low-energy bands of the anisotropy model have a topological character, which is captured by the Berry curvature dipole \cite{BCD_mismatch, BCD_strain}. The Berry curvature dipole (BCD) is defined as $\Lambda_{i}=\sum_n \int_{MBZ} \frac{d^2 \mathbf{k}}{(2\pi)^2}\Omega^n \frac{\partial E^n(\mathbf{k})}{\hbar \partial k_i}\frac{\partial f(E^n(\mathbf{k}))}{\partial E^n(\mathbf{k})}$ \cite{BCD_strain}, where $n$ labels the band, $f(E)$ is the Fermi-Dirac distribution function, and $\Omega$ is the Berry curvature. In order for the BCD to be nonzero, one must have nonzero Berry curvature, broken inversion symmetry, and broken rotational symmetry. Despite the Chern number being zero due to time-reversal symmetry, the anisotropy model satisfies the requirements for nonzero BCD: there is nonzero Berry curvature arising from the massive Dirac cones, the inversion symmetry is broken by the TMD monolayer, and the $C_3$ rotational symmetry is broken by the Moir\'e potential. The Berry curvature near the massive Dirac cones scales with the Dirac mass like $m^{-2}$, and thus the BCD can be significantly enhanced by a negative $w_2$ which reduces the Dirac mass. Throughout the realistic parameter regimes explored in our simulations, we find that when the Fermi level is near the bottom/top of the first conduction/valence band, the BCD is generically present and on the order of $0.1 - 1 \mathrm{\r{A}}$ when $w_2=0$. The BCD is tied to the nonlinear hall effect \cite{nonlinear_hall}, which can be used to experimentally probe it.

\section{$C_{3}$ Symmetric Model}
\label{sec:c3_model}

\begin{figure*}[t]
    \centering
    \hspace*{-3cm}\includegraphics[height=10cm, width=22cm]{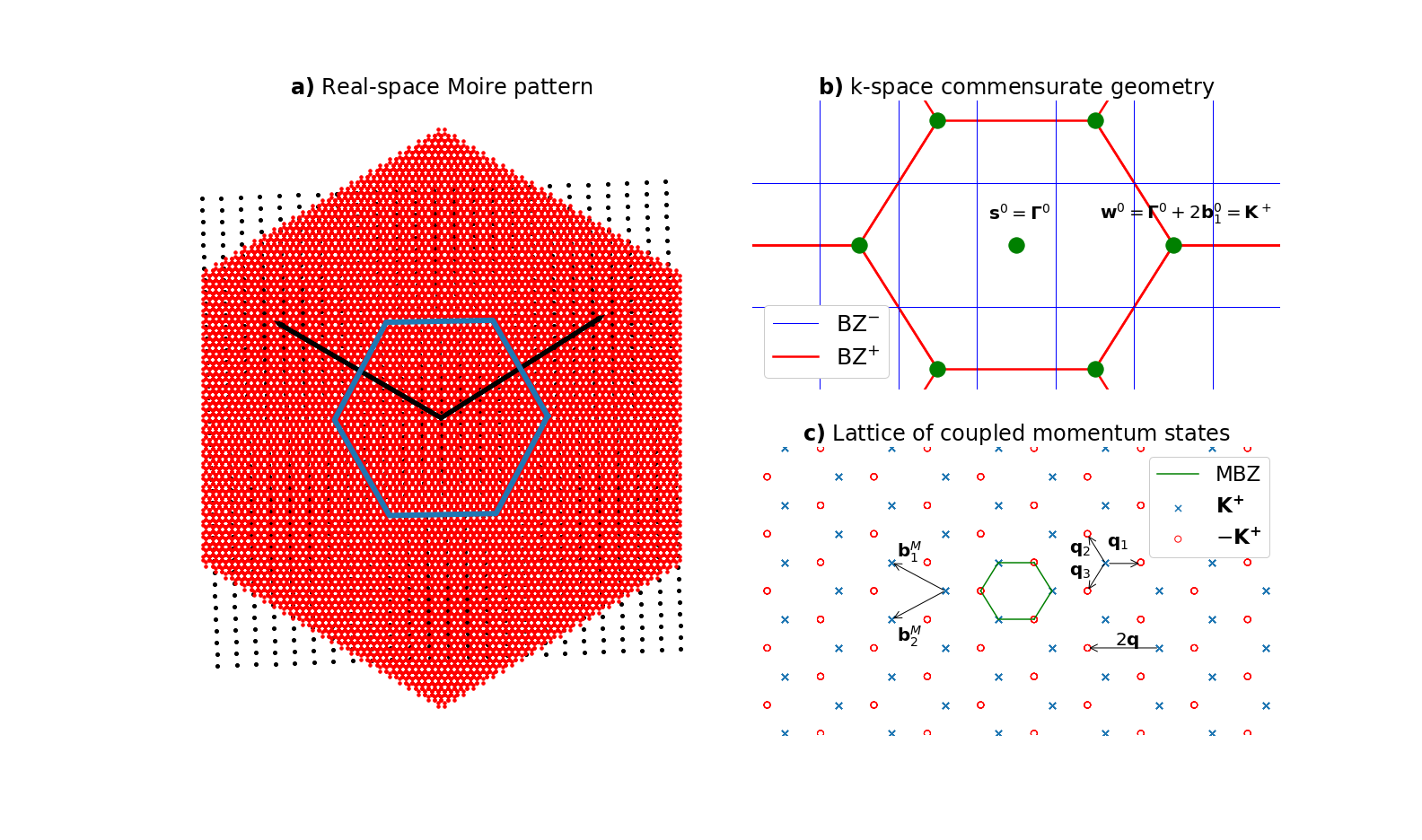}
    \caption{$\mathbf{C_3}$ \textbf{Model Geometry} \textbf{a)} Moir\'e pattern, $\theta=2\degree$ and $\epsilon=0$. Blue hexagon denotes Moir\'e unit cell, and black arrows denote Moir\'e primitive vectors. \textbf{b)} Momentum space commensurate geometry. Low-energy Moir\'e couplings are generated by substrate states near $\boldsymbol{\Gamma}^-$. \textbf{c)} Lattice of momentum states coupled by Moir\'e potential. The states form a honeycomb lattice, like in the Bistritzer--Macdonald model of twisted bilayer graphene \cite{bm_graphene}.}
    \label{fig:c3_geometry}
\end{figure*}

\begin{figure*}[t]
    \centering
    \hspace*{-2.2cm}\includegraphics[height=6cm, width=22cm]{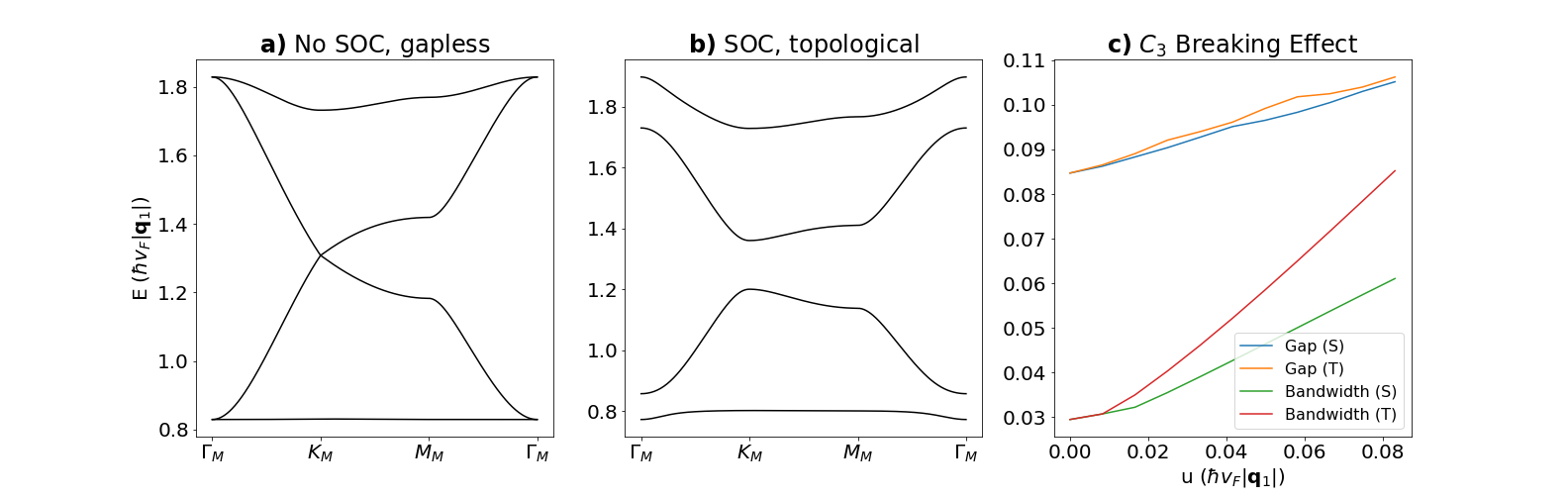}
    \caption{\textbf{Topological Flat Bands in} $\mathbf{C_3}$ \textbf{Model} Bandstructure hosting a flat band \textbf{a)} without and \textbf{b)} with spin-orbit coupling (SOC). We show only the four bands nearest charge neutrality. The SOC gaps out the bands, giving the bottom (top) band spin Chern number $C_{\uparrow} = -1$ $(C_{\uparrow} =+1)$. The corresponding spin down bands have opposite Chern numbers, giving rise to a quantum spin hall effect. \textbf{c)} Effect of $C_3$ breaking perturbations on the energy gap and the bandwidth of the bottom band. For comparison, a typical (non-flat) bandwidth is on the order of $0.3$. We consider perturbations on intravalley (S) and intervalley (T) hoppings. The strength of the perturbations are denoted by $u$.}
    \label{fig:c3_breaking}
\end{figure*}

For the $C_3$ symmetric model, we consider the case of a graphene monolayer as the active layer, and a rectangular substrate geometry given by $\mathbf{b}_1^0 = (\frac{2\pi}{3a}, 0)$, $\mathbf{b}_1^0 = (0, \frac{2\pi}{a\sqrt{3}})$, $\mathbf{b}_i^- = R_{\theta}e^{\epsilon}\mathbf{b}_i^0$. In this case, $\mathbf{b}_i^c = \mathbf{b}_i^+$, and therefore the Moir\'e BZ is hexagonal. The geometry is pictured in Figure \ref{fig:c3_geometry}. Here, $S^0 = \{\boldsymbol{\Gamma}^0\}$. Thus, the graphene states in either valley may tunnel into the substrate layer at k-points near $\boldsymbol{\Gamma}^-$ before tunneling back, with a momentum offset of $2\mathbf{q} = \frac{2}{3}(\mathbf{b}_1^M + \mathbf{b}_2^M)$ to switch valleys. We define $\mathbf{q}_1 = 2\mathbf{q}-\mathbf{b}_1^M-\mathbf{b}_2^M$, $\mathbf{q}_2 = 2\mathbf{q}-\mathbf{b}_2^M$, $\mathbf{q}_3 = 2\mathbf{q}-\mathbf{b}_1^M$ as the vectors which connect a momentum state in one valley, $\mathbf{K}^+ + \mathbf{k}$, with the nearest neighbor momentum states that it can tunnel to in the opposite valley, $-\mathbf{K}^+ + \mathbf{k}+\mathbf{q}_i$ (see Figure \ref{fig:c3_geometry}\textbf{d}). The tunneling directions are $C_3$ symmetric, in the sense that $\mathbf{q}_{i+1}= R_{2\pi/3}\mathbf{q}_i$. Because of this fact, the continuum model for this configuration inherits the $C_3$ symmetry of graphene under some weak assumptions, despite the explicit $C_3$ symmetry breaking in the substrate. Firstly we assume that the real-space microscopic tunneling amplitude $\hat{t}^{\alpha\beta}(\mathbf{\mathbf{r}})$ depends only on the norm of its argument, with no angular dependence. This is a reasonable assumption especially for orbitals without azimuth angle dependence e.g. $s$, $p_z$, and $d_{z^2}$ orbitals. If we also assume that there is only one active orbital near the graphene Fermi energy, located at $\boldsymbol{\tau}^- = (0, 0)$, then the tunneling amplitudes (Equations \ref{intravalley_pot} and \ref{intervalley_pot}) are $C_3$ symmetric, because the tunneling directions and Moir\'e reciprocal lattice are $C_3$ symmetric. If we collect the graphene Bloch states in each valley on each orbital into a four-component vector, $\ket{\mathbf{k}} = (\ket{\mathbf{K}^+ + \mathbf{k}, A}, \ket{\mathbf{K}^+ + \mathbf{k}, B}, \ket{-\mathbf{K}^+ + \mathbf{k}, A}, \ket{-\mathbf{K}^+ + \mathbf{k}, B})$ then the $C_3$ symmetry acts as $C_3\ket{\mathbf{k}} = \ket{R_{2\pi/3}\mathbf{k}}e^{i\frac{2\pi}{3}\sigma_z \otimes \sigma_z}$. The model also possesses mirror symmetries $M_x$ and $M_y$, and if we assume that the substrate possesses the spinful anti-unitary time reversal symmetry $\tau$ with $\tau^2 = -1$, this time reversal symmetry is inherited by the Moir\'e model as well. As both spin species contribute to the low energy sector of the theory, we collect both the spin and orbital degrees of freedom into the creation operators $a^\dag_\mathbf{k}, b^\dag_\mathbf{k}$, so that the hopping matrices are 4x4 complex matrices (the Pauli $\sigma$ matrices appearing in the dispersion still acts only on the orbital degrees of freedom). Again keeping only the few lowest order tunneling matrices and defining $\mathbf{k}_{nm}^{\pm} = \mathbf{k} + n\mathbf{b}_1^M + m\mathbf{b}_2^M \pm \frac{1}{2}\mathbf{q}_1$, the Hamiltonian is given by

\begin{gather}
\begin{split}
\label{c3_model}
&H = \frac{\Omega_M}{(2\pi)^2}\int_{MBZ} d^2\mathbf{k} H(\mathbf{k}) \\&\\
H(\mathbf{k}) = &\sum_{n,m} [a^\dag_{\mathbf{k}_{nm}^+}(\hbar v_{F}\boldsymbol{\sigma}\cdot\mathbf{k}_{nm}^+ + S_0^+) a_{\mathbf{k}_{nm}^+} \\
 + b^\dag_{\mathbf{k}_{nm}^-}&(-\hbar v_{F}\boldsymbol{\sigma}^*\cdot\mathbf{k}_{nm}^- + S_0^-)  b_{\mathbf{k}_{nm}^-} \\
+ \{a^\dag_{\mathbf{k}_{(n+1)m}^+} S_1^+ a_{\mathbf{k}_{nm}^+}&
+ a^\dag_{\mathbf{k}_{n(m+1)}^+} S_2^+ a_{\mathbf{k}_{nm}^+} + a^\dag_{\mathbf{k}_{(n+1)(m-1)}^+} S_3^+ a_{\mathbf{k}_{nm}^+} \\
+ b^\dag_{\mathbf{k}_{(n-1)m}^-} S_1^- b_{\mathbf{k}_{nm}^-}& + b^\dag_{\mathbf{k}_{n(m-1)}^-} S_2^- b_{\mathbf{k}_{nm}^-} + b^\dag_{\mathbf{k}_{(n-1)(m+1)}^-} S_3^- b_{\mathbf{k}_{nm}^-}  \\
+ b^\dag_{\mathbf{k}_{nm}^-} T_1 a_{\mathbf{k}_{nm}^+}& + b^\dag_{\mathbf{k}_{(n-1)m}^-} T_2 a_{\mathbf{k}_{nm}^+} + 
b^\dag_{\mathbf{k}_{n(m-1)}^-} T_3 a_{\mathbf{k}_{nm}^+} \\ 
+ b^\dag_{\mathbf{k}_{(n-1)(m-1)}^-} &T_4 a_{\mathbf{k}_{nm}^+} + b^\dag_{\mathbf{k}_{(n+1)(m-1)}^-} T_5 a_{\mathbf{k}_{nm}^+} \\
+ b^\dag&_{\mathbf{k}_{(n-1)(m+1)}^-} T_6 a_{\mathbf{k}_{nm}^+} + \mathrm{h.c.}\}]
\end{split}
\end{gather}

\noindent Here, $T_{1/2/3}$ describe (intervalley) nearest neighbor hopping, $S_{1/2/3}^\pm$ describe (intravalley) next neighbor hopping, and $T_{4/5/6}$ denote (intervalley) third neighbor hopping in momentum space. Unlike the spinless case considered above, the intravalley hopping matrices in each valley $S_i^\pm$ are not just related by complex conjugation, however they are still related by spinful time reversal symmetry. This model is equivalent to a model that was studied extensively in Ref.~\cite{Kagome_flat_bands} (see Appendix \ref{app:c3_map}), in which it arose as a result of graphene on top of a hexagonal substrate geometry. They also discuss the constraints on the hopping matrices imposed by the symmetries, and we refer the interested reader to their work for the details. In their case, there is a further decomposition of the Moir\'e potentials in terms of the corepresentations carried by the substrate states under the symmetries. In our case, where the substrate does not possess $C_3$ symmetry, no such decomposition exists. They report an interesting phase diagram featuring flat spin Chern bands that can be accurately captured by tight-binding models on Kagome and honeycomb lattices \cite{kagome_tb, honeycomb_tb, honeycomb_tb_moire} in the case of the substrate states carrying a single 2D spinless coirrep of the space group $P6mm1'$. These settings are promising candidates for anomalous quantum hall states, quantum spin liquids, Mott insulators, etc. \cite{kagome_magnets, kagome_qah, kagome_spin_liquid, honeycomb_tb_moire} The interesting parameter regimes considered in Ref.~\cite{Kagome_flat_bands} are accessible in our model as well, though without the help of the corepresentation decomposition, it may require more fine tuning to realize those phases in our setting.

We can straightforwardly estimate the sources of $C_3$ symmetry breaking in our model. Firstly, the Moir\'e potential couples the K/K' valleys to other high energy states in the graphene layer that we projected away in deriving the continuum model, and these couplings are not $C_3$ symmetric. These states include k-points near $\boldsymbol{\Gamma}^+$ and $\pm\frac{1}{2}\mathbf{K}^+$ and have energies on the order of $U^+ \approx 5 \mathrm{eV}$. Denoting the energy of the substrate states near $\boldsymbol{\Gamma}^-$ closest to the graphene Fermi level by $U^-$, these states will lead to perturbations of the low-energy continuum model stemming from fourth and higher order processes, e.g. $\mathbf{K}^+ \rightarrow \boldsymbol{\Gamma}^- \rightarrow \frac{1}{2}\mathbf{K}^+ \rightarrow \boldsymbol{\Gamma}^- \rightarrow -\mathbf{K}^+$, which are weaker than the $C_3$ symmetric processes (of strength $\frac{\hat{t}^2}{U^-}$) by a factor of $\frac{\hat{t}^2}{U^+ U^-}$. Next, the dependence of the substrate Hamiltonian $H^-$ around the $\boldsymbol{\Gamma}^-$ point on the wavevector $\mathbf{k}$, which we neglect at lowest order, also breaks the $C_3$ symmetry. If we assume a linear dispersion of velocity $v^-$ in the substrate around the $\boldsymbol{\Gamma}^-$ point, then this will introduce perturbations to the continuum model that are weaker than the $C_3$ processes by a factor of $\frac{\hbar v^- |\mathbf{q}|}{U^-}$.

To investigate the effect of the $C_3$ symmetry breaking perturbations on the phases of interest, we tune the model Equation \ref{c3_model} to a particular parameter regime that hosts topological (i.e., nonzero spin Chern number) flat bands, and consider the evolution of the band structure as a $C_3$ breaking perturbation is turned on. Generically the spin Chern number is stable so long as the gap is not closed, but the bandwidth will vary continuously as a function of the perturbation strength. The gap is controlled by the strength of the spin-orbit coupling (SOC). In Figure \ref{fig:c3_breaking} we display a typical band structure of the $C_3$ symmetric model hosting a topological flat band. We then add various $C_3$ breaking perturbations to the hopping matrices, and track the bandwidth of the topological flat band and the gap to the next closest band as a function of the perturbation strength. The hoppings we consider preserve the z component of spin, and we plot only the bands for spin up electrons. The chosen range for the perturbation strength is justified in Appendix \ref{app:c3_map} by estimating the typical strength of the perturbations in a realistic material. We find that the gap remains robust and the bandwidth remains suppressed compared to a typical non-flat Moir\'e band.

\section{Conclusions}

In this paper, we presented a generic construction of Moir\'e ``coupled-valley'' models, in which the K and K' valleys of a honeycomb active layer are coupled via second order hopping processes through a substrate. We derived the continuum model and the Moir\'e potentials from the microscopic tunneling amplitudes, which can be calculated directly from the orbitals and atomic potentials of each layer. We then identified two symmetry-mismatched geometries which give rise to interesting Moir\'e physics. 

The first model displays an emergent time-reversal symmetry, and gives rise to a strongly anisotropic regime in which a Sliding Luttinger Liquid phase exists, displaying a quasi-1D character. The origin of the anisotropy in this model is distinct from other anisotropic Moir\'e models considered in the literature, as in this case the active monolayer is perfectly isotropic before the Moir\'e potential is turned on, and the anisotropy is fundamentally due to the specific Brillouin Zone folding induced by the Moir\'e potential. Therefore this setting does not fall under the purview of various Moir\'e models considered previously in which the Moir\'e potential serves to enhance the intrinsic anisotropy of the monolayer \cite{Coupled_wire, anisotropy-moire}. The bands in this model also possess a nonzero Berry curvature dipole, owing to the broken inversion and rotational symmetries. 

The second model displays an emergent three-fold rotational symmetry which is not present in the substrate. The continuum model reduces to one considered previously in Ref.~\cite{Kagome_flat_bands}, which was shown to display a rich zoo of topological flat bands, including Kagome and honeycomb flat bands. This mechanism of emergent symmetry is somewhat generic, in that the coupled-valley continuum model can boast a discrete symmetry of the active layer even when that symmetry is explicitly broken by the substrate, so long as the lattice of low-energy states in both valleys coupled by the Moir\'e potential retains that symmetry. We also showed that the $C_3$ breaking perturbations on the model do not necessarily destroy the topological flat bands.

\section{Acknowledgements}

AS is grateful for the guidance offered by Chao-Ming Jian while studying at Cornell University, and for the many fruitful conversations had with him. AS also acknowledges the Cornell University Nexus Scholars program, which partially funded him during the time period that this work was conducted. 


\bibliography{bibliography}

\appendix

\section{Coupled-Valley Continuum Model Construction Recipe}
\label{derivation}
We now present a method to construct a general coupled-valley continuum model. 

Suppose that the single-particle Hilbert space on each layer can be expanded in an orthonormal tight-binding basis, $\ket{l, \mathbf{R}_j, \alpha}$, where $l=(\pm)$ denotes the layer, $\mathbf{R}_j$ denotes the jth unit cell in the corresponding layer, and $\alpha$ indexes the orbitals in the corresponding layer. Under the two-center approximation, the hopping element between the two layers depends only on the real-space separation of the tight-binding basis states and their orbital character, and therefore takes the form $$\braket{-, \mathbf{R}_j, \beta | H | +, \mathbf{R_i}, \alpha} = t^{\alpha\beta}((\mathbf{R}_i+\boldsymbol{\tau}_\alpha^+) - (\mathbf{R}_j+\boldsymbol{\tau}_\beta^-)).$$ Fourier transforming, we denote the Bloch states $\ket{l, \mathbf{k}, \alpha} = \frac{1}{\sqrt{|BZ_l|}}\sum_j \ket{l, \mathbf{R}_j, \alpha} e^{-i\mathbf{k}\cdot(\mathbf{R_j}+\boldsymbol\tau_\alpha^l)}$, and we can write the tunneling amplitude as \cite{bm_graphene, Hypermagic_flat_bands} 

\begin{align}
\begin{split}
\label{mom_cons}
&\braket{-, \mathbf{k'}, \beta | H | +, \mathbf{k}, \alpha} = \\
&\frac{1}{\sqrt{|\Omega_+||\Omega_-|}}\sum_{\mathbf{G^+}, \mathbf{G^-}}\hat{t}^{\alpha\beta}(\mathbf{k}+\mathbf{G^+})e^{i(\mathbf{G}^+\cdot \boldsymbol\tau_\alpha^+ - \mathbf{G}^-\cdot \boldsymbol\tau_\beta^-)}\\
&\delta^2((\mathbf{k}+\mathbf{G}^+) - (\mathbf{k'}+\mathbf{G}^-))
\end{split}
\end{align}

\noindent where $\mathbf{G}^{l}\in \mathrm{Span}(\mathbf{b}_{1}^l, \mathbf{b}_{2}^l)$ is a reciprocal lattice vector in layer $l$, and $|\Omega_l|$ denotes the primitive unit cell area in layer $l$. 

We can glean several important pieces of information from this formula. Firstly, the delta function enforces conservation of crystal momentum, so that tunneling between states in the two layers is only allowed if their crystal momenta are equivalent modulo a sum of reciprocal lattice vectors in each layer. Recall we are ultimately interested in tunneling of states near $\pm\mathbf{K}^+$. The consequences of conservation of crystal momentum on the allowed tunnelings is best appreciated by considering the commensurate geometry. Adapting the notation from Ref.~\cite{Kagome_flat_bands}, we define $+S^+ = (\mathbf{K^+}+\mathrm{Span}(\mathbf{b}_1^+, \mathbf{b}_2^+, \mathbf{b}_1^0, \mathbf{b}_2^0)) \cap BZ^+$, $+\tilde{S}^0 = (\mathbf{K^+}+\mathrm{Span}(\mathbf{b}_1^+, \mathbf{b}_2^+, \mathbf{b}_1^0, \mathbf{b}_2^0)) \cap BZ^0$. In the commensurate case, $\mathbf{K}^+$ can tunnel to any state in the bottom layer with crystal momentum given by an element of $+\tilde{S}^0$. Conversely, any state in the bottom layer with crystal momentum given by an element of $+\tilde{S}^0$ can tunnel to any state in the top layer with crystal momentum given by an element of $+S^+$. Thus, second order tunneling processes will mix $\mathbf{K^+}$ with all of the other momenta states in $+S^+$. We likewise define $-S^+ = (-\mathbf{K^+}+\mathrm{Span}(\mathbf{b}_1^+, \mathbf{b}_2^+, \mathbf{b}_1^0, \mathbf{b}_2^0)) \cap BZ^+$, $-\tilde{S}^0 = (-\mathbf{K^+}+\mathrm{Span}(\mathbf{b}_1^+, \mathbf{b}_2^+, \mathbf{b}_1^0, \mathbf{b}_2^0)) \cap BZ^0$ as the states that $-\mathbf{K^+}$ can tunnel into. It is obvious that the elements of $-\tilde{S}^0$ are the same as the elements of $\tilde{S}^0$ multiplied by $-1$. Therefore, if the set $S^0 = \{\mathbf{\tilde{s}}^0 \in \tilde{S^0} : -\mathbf{\tilde{s}}^0 \in (\mathbf{\tilde{s}}^0 +\mathrm{Span}(\mathbf{b}_1^0, \mathbf{b}_2^0))\}$ is nonempty, then the tunneling process $\mathbf{K^+} \rightarrow \mathbf{s}^0\in S^0 \rightarrow \mathbf{-K^+}$ is allowed, giving us an intervalley coupling. This also implies that $+S^+ = -S^+$, or in other words, that the two valleys are exactly folded into each other in the commensurate case. $S^0$ being nonempty is therefore a sufficient constraint on the commensurate substrate geometry to get a coupled-valley model.

Secondly, it is important to note that in Eq.~\ref{mom_cons}, the tunneling amplitude only depends on the bottom layer geometry through the crystal momentum conservation equation and the geometrical factor in the exponent. The functions $\hat{t}^{\alpha\beta}(\mathbf{q})$ depend only on the orbital character of the wavefunctions in each layer and the atomic potentials. Because of this, even when the substrate explicitly breaks a discrete (e.g. rotational) symmetry of the active layer, this symmetry may be preserved in the coupled-valley continuum model. This situation arises in the two explicit models we consider in Sections \ref{sec:anisotropy_model} and \ref{sec:c3_model}.

Assuming $S^0$ is nonempty, we can now use the crystal momentum conservation Eq.~\ref{mom_cons} to derive the Moir\'e potentials. Second order processes to $\pm\tilde{S}^0$ can generate intravalley hopping processes, while second order processes to $S^0$ can generate both intravalley and intervalley hopping processes. For ease of notation we assume that $S^0 =  \pm\tilde{S}^0$, so that every allowed hopping process generates both an intravalley and an intervalley coupling. If this is not the case, one can straightforwardly add back in the intravalley couplings generated by the sets $\pm \tilde{S}^0 \backslash S^0$. Consider one k-point $\mathbf{s}_i^0 \in S^0$. Let $\mathbf{w}_i^0$ be an element of $(\mathbf{s}_i^0+\mathrm{Span}(\mathbf{b}_1^0, \mathbf{b}_2^0))\cap(\mathbf{K^+}+\mathrm{Span}(\mathbf{b}_1^+, \mathbf{b}_2^+))$ with minimal norm. Note that these points exist in the commensurate substrate configuration, so the physical k-points they correspond to in the substrate are $\mathbf{s}_i^- = R_{\theta}e^{-\epsilon}\mathbf{s}_i^0$, $\mathbf{w}_i^- = R_{\theta}e^{-\epsilon}\mathbf{w}_i^0$. Suppose that the substrate Hamiltonian is gapped at $\mathbf{s}_i^-$ (relative to the active layer Fermi level), and denoted by $H_i^- = H^-(\mathbf{s}_i^-)$. Let $\mathbf{q}_i = \mathbf{w}_i^- - \mathbf{w}_i^0 = -(1-R_{\theta}e^{-\epsilon})\mathbf{w}_i^0$. Then, it is straightforward to show that

\begin{align}
\begin{split}
    &\braket{-, \mathbf{s}_i^-+\mathbf{k'}, \beta|H|+, \mathbf{K^+}+\mathbf{k}, \alpha} = \\
    &\frac{1}{\sqrt{|\Omega_+||\Omega_-|}}\sum_{n,m}\hat{t}^{\alpha\beta}(\mathbf{w}_i^0+n\mathbf{b}_1^c+m\mathbf{b}_2^c)\\
    &e^{i\phi_{nmi}^{\alpha\beta}}\delta^2(\mathbf{k'} - (\mathbf{k}+\mathbf{q}_i+n\mathbf{b}_1^M+m\mathbf{b}_2^M)),\\\\
    &\braket{-, \mathbf{s}_i^-+\mathbf{k'}, \beta|H|+, -\mathbf{K^+}+\mathbf{k}, \alpha} = \\
    &\braket{-, -\mathbf{s}_i^-+\mathbf{k'}, \beta|H|+, -\mathbf{K^+}+\mathbf{k}, \alpha} = \\
    &\frac{1}{\sqrt{|\Omega_+||\Omega_-|}}\sum_{n,m}\hat{t}^{\alpha\beta}(-\mathbf{w}_i^0-n\mathbf{b}_1^c-m\mathbf{b}_2^c)\\
    &e^{-i\phi_{nmi}^{\alpha\beta}}\delta^2(\mathbf{k'} - (\mathbf{k}-\mathbf{q}_i-n\mathbf{b}_1^M-m\mathbf{b}_2^M)).
\end{split}
\end{align}

\noindent where we utilized the fact that $\mathbf{w}_i^0+n\mathbf{b}_1^c+m\mathbf{b}_2^c + \mathbf{k}+\mathbf{q}_i \approx \mathbf{w}_i^0+n\mathbf{b}_1^c+m\mathbf{b}_2^c$ in the argument of $\hat{t}^{\alpha\beta}$ due to $\mathbf{k}$ and $\mathbf{q}_i$ being small. From the formulae, it is clear that the intervalley hopping processes $\mathbf{K}^+ + \mathbf{k} \rightarrow \mathbf{s}_i^- + \mathbf{k}+\mathbf{q}_i + a\mathbf{b}_1^M + b\mathbf{b}_2^M \rightarrow -\mathbf{K}^+ + \mathbf{k} + 2\mathbf{q}_i + n\mathbf{b}_1^M + m\mathbf{b}_2^M$ are allowed, while all other intervalley momentum transfers are forbidden. Likewise, the intravalley hopping processes $\pm\mathbf{K}^+ + \mathbf{k} \rightarrow \mathbf{s}_i^- + \mathbf{k}\pm\mathbf{q}_i + a\mathbf{b}_1^M + b\mathbf{b}_2^M \rightarrow \pm\mathbf{K}^+ + \mathbf{k} + n\mathbf{b}_1^M + m\mathbf{b}_2^M$ are allowed. Therefore, intravalley hopping processes are given by hoppings on the Moir\'e reciprocal lattice, while intervalley hopping processes are given by hoppings on the Moir\'e reciprocal lattice with a momentum offset given by $2\mathbf{q}_i$ to account for changing the valley. 

To visualize the various momenta described above, consider the geometry pictured in Figure \ref{fig:anisotropy_geometry}, which was analyzed in Section \ref{sec:anisotropy_model}. As shown in Figure \ref{fig:anisotropy_geometry}c, in the commensurate geometry, $\mathbf{s}_1^0=\boldsymbol{\Gamma}^0$ and $\mathbf{s}_2^0=\mathbf{M}^0$ are the only points in the substrate Brillouin Zone equivalent to $\mathbf{K}^+$ modulo reciprocal lattice vectors. Since $\boldsymbol{\Gamma}^0 = -\boldsymbol{\Gamma}^0$ and $\mathbf{M}^0 = -\mathbf{M}^0 + \mathbf{b}_1^0$, both points generate both intervalley and intravalley couplings. The lowest norm points where $\boldsymbol{\Gamma}^0$ and $\mathbf{M}^0$ (plus substrate reciprocal lattice vectors) intersect $\mathbf{K}^+$ (plus active layer reciprocal lattice vectors) are $\mathbf{w}_1^0 = \mathbf{K}^+$ and $\mathbf{w}_2^0 = \mathbf{M}^0$. In the incommensurate case, the mismatch of these two points with their incommensurate counterparts generates the vectors $\mathbf{q}_1 = -(1-R_{\theta}e^{-\epsilon})\mathbf{K}^+$, $\mathbf{q}_2 = -(1-R_{\theta}e^{-\epsilon})\mathbf{M}^0$, which generate the lattice mismatch on the Moir\'e scale shown in Figure \ref{fig:anisotropy_geometry}d. 

From here, it is straightforward to use the Schrieffer-Wolff perturbation theory to derive the result for the continuum model displayed in the main text. In deriving this model we have also projected out all of the active layer momentum states near points in $\pm S^+$ besides the low-energy states near $\pm\mathbf{K^+}$. In principle these states will lead to perturbations of the low-energy model stemming from weak fourth-order and higher tunneling processes.

\section{Parameters for the Anisotropy Model}
\label{app:anisotropy_params}

In this section we briefly state the explicit forms of all of the tunneling matrices in the anisotropy model based on the discrete symmetry constraints ($M_x$ and $\tau$). Here, $w_{0-24}$ are real parameters, and the Pauli matrices act on the sublattice index.

\begin{align}
\begin{split}
    &S_0 = w_0 \sigma_0 + w_1 \sigma_x + w_2 \sigma_z \\
    &T_1 = w_{3}\sigma_0 + w_{4}\sigma_x + w_{5}\sigma_z \\
    &T_2 = w_{6}\sigma_0 + w_{7}\sigma_x + w_{8}\sigma_z \\
    &S_1 = w_{9} \sigma_0 + w_{10} \sigma_x + i w_{11} \sigma_y + w_{12} \sigma_z \\
    &S_2 = (w_{13}+i w_{14})\sigma_0 + (w_{15} + i w_{16})\sigma_x + (w_{17}+i w_{18})\sigma_z \\
    &T_3 = (w_{19}+i w_{20})\sigma_0 + (w_{21}+i w_{22})\sigma_x + (w_{23}+i w_{24})\sigma_z \\
    &T_4 = (w_{19}-i w_{20})\sigma_0 + (w_{21}-i w_{22})\sigma_x + (w_{23}-i w_{24})\sigma_z
\end{split}
\end{align}

In Figures \ref{fig:anisotropy_bs}, \ref{fig:anisotropy_fs}\textbf{d)}, and \ref{fig:anisotropy_fs}\textbf{e)}, $w_3 = 70\mathrm{meV}$, $w_6 = 40\mathrm{meV}$, $w_9 = 20 \mathrm{meV}$, and $w_{13} = 40\mathrm{meV}$, $w_1$ varies, and all other parameters are zero. In Figures \ref{fig:anisotropy_fs}\textbf{a)}, \ref{fig:anisotropy_fs}\textbf{b)}, we set all couplings equal to zero except $w_{1}$ and $w_3$, the latter of which corresponds the label $|T_1| = |w_3|$ shown in the figures. This was done to speed up the computation time, allowing us to consider only the smallest two momentum states $\mathbf{k}_{00}^{\pm}$. Such a procedure is justified because this minimal two-k-point model approximates the energetics of the first conduction band well, except for near the MBZ edges where it differs from the full model slightly due to hybridization with other k-point states. This error near the edges is insignificant compared to the bandwidths, and can thus be neglected.

\section{Parameters for the $C_3$ Model}
\label{app:c3_map}

As stated in the main text, our model (in the absence of the $C_3$ breaking perturbations) is exactly equivalent to the coupled-valley model considered in Ref.~\cite{Kagome_flat_bands}. By applying the unitary transformation $b_\mathbf{k}^\dag = \tilde{b}_\mathbf{k}^\dag \sigma_y$, $b_\mathbf{k} = \sigma_y \tilde{b}_\mathbf{k}$, rotating to a Moir\'e coordinate system $(\hat{\mathbf{x}}^M, \hat{\mathbf{y}}^M)$ so that $\mathbf{q}_1 = |\mathbf{q}_1|\hat{\mathbf{y}}^M$, and moving to real space, the two models are manifestly the same. To investigate the effect of $C_3$ symmetry breaking on the bandwidth of the flat Chern bands hosted by the model, we choose the same parameter values as in Figure 3\textbf{a)} of their text, and add a weak $C_3$ symmetry breaking potential that preserves the other discrete symmetries of the model. For a rough order of magnitude estimate of how large the $C_3$ symmetry breaking potential should be, we assume the Moir\'e lattice constant is on the order of 10 \r{A}. Then, the characteristic graphene monolayer kinetic energy scale $E_{\mathrm{scale}} = \hbar v_F |\mathbf{q}_1|$ and the hopping matrices without the $C_3$ breaking potential and without spin-orbit coupling are of order $500 \mathrm{meV}$, while the modifications to the hopping matrices from the spin-orbit coupling (SOC) are of order $100 \mathrm{meV}$. More precisely, for $|\mathbf{q}_1|=\frac{1}{10} \mathrm{\r{A}}^{-1}$, $E_{\mathrm{scale}}=601\mathrm{meV}$, and the largest contribution to the tunneling matrices is $409\mathrm{meV}$, while the largest contribution to the tunneling matrices from the SOC is $152\mathrm{meV}$. Assuming the substrate energy relative to the graphene Fermi level is of order $U^- \sim 5\mathrm{eV}$, this sets the characteristic energy scale of the tunneling amplitude $|\hat{t}(\mathbf{q}_1)|$, as $\frac{|\hat{t}(\mathbf{q}_1)|^2}{U^-} \sim 500\mathrm{meV}$. This gives a characteristic strength for the $C_3$ breaking perturbations as $\frac{\hat{t}^4}{(U^-)^2 U^+} \sim 5\mathrm{meV}$ and $\frac{\hat{t}^2\hbar v^- |\mathbf{q}|}{(U^-)^2} \sim 25\mathrm{meV}$ where we assumed that $v^-$ is upper bounded by the graphene Fermi velocity, which will generally be an overestimation for typical substrates. As a conservative estimate, we therefore allow the $C_3$ symmetry breaking perturbation strength to take a maximum value of $50\mathrm{meV}$, or roughly $30\%$ of the SOC strength. We consider two kinds of $C_3$ breaking perturbations: firstly, in the notation of Ref.~\cite{Kagome_flat_bands}, we perturb $w_{\sqrt{3}, x} \rightarrow w_{\sqrt{3}, x}+u$ away from it's original value for only $S_{s,\eta,\gamma(\mathbf{q}_2-\mathbf{q}_3)}$, while leaving $S_{s,\eta,\gamma(\mathbf{q}_3-\mathbf{q}_1)}$, $S_{s,\eta,\gamma(\mathbf{q}_1-\mathbf{q}_2)}$ unchanged. The second perturbation we consider is $w_{1,y}\rightarrow w_{1,y}+u$ for only $T_{s,\mathbf{q}_1}$, while leaving $T_{s,\mathbf{q}_2}$, $T_{s,\mathbf{q}_3}$ unchanged. Thus we consider the effect of $C_3$ breaking perturbations on both intravalley and intervalley hopping. We display results in both cases for $u\geq 0$, but find the results are very similar for $u<0$.

\end{document}